\documentstyle[aps, preprint]{revtex}
\begin{document}
\tightenlines
\preprint{}
\title{Possibility of extracting the weak phase $\gamma$ from $\Lambda_b \to
\Lambda D^0$ decays}
\author{A. K. Giri$^1$, R. Mohanta$^2$ and M. P. Khanna$^1$ }
\address{ $^1$ Physics Department, Panjab University, Chandigarh-160 014,
India}
\address{ $^2$ School of Physics, University of Hyderabad,
Hyderabad-500 046, India }
\maketitle
\begin{abstract}
We explore the possibility of extracting the weak phase $\gamma$ from
pure tree decays $\Lambda_b
\to \Lambda (D^0, \overline{D^0}, D^0_{CP})$ in a model independent way.
The CP violating weak phase $\gamma$ can
be determined cleanly, without any hadronic uncertainties,
as these decay modes are free from the 
penguin pollutions. Furthermore, neither tagging nor time dependent
studies are required to extract the angle $\gamma $ with these
modes.
\end{abstract}
\pacs{PACS Nos. : 12.15.Hh, 13.30.Eg, 14.20.Mr}

\section{Introduction}

CP violation still remains one of the unsolved problems till date in
particle physics \cite {kog89,bigi00,lav00}, even after almost four decades
of its discovery in 1964 in neutral $K$ meson system.
Since then various attempts have been made
by theorists and experimentalists to understand it but without much
success.  The Standard Model (SM) provides a simple description
of this phenomenon through the complex CKM matrix \cite{ref1}.
Decays of $B$ mesons provide a rich ground for investigating CP
violation \cite{buras98,quin98}. They allow stringent tests both for
the SM and for studies of new sources of this effect. Within the SM, CP
violation is often characterized by the so called unitarity triangle
\cite{chau94}. Detection of CP violation and the accurate determination
of the unitarity triangle are the major goals of experimental $B$ physics
\cite{stone94}.
Although the exact cause of its origin
is not yet understood completely but the recent results
from SLAC and KEK $B$-factories, the first evidence of large CP
violation in $B$-systems \cite{belle01,babar01}, have
necessitated renewed interests in understanding the nature of CP violation
within the framework of Standard Model (SM), instead of looking beyond it.
Therefore, it is really high time in particle physics in general and
$B$-physics in particular since CP violation is interlinked to many
problems in particle physics. It can also give us a possible explanation
for the baryon
asymmetry of the universe. The general expectation is to check all possible
ways to find out the root cause of it and
explore as many decay modes as possible to arrive at a decisive conclusion.
We now believe that the CKM explanation
of SM (which explains the CP violation in K-systems), 
may also explain the mechanism of CP violation in $B$-systems.
The knowledge gained in exploring various scenarios in
$B$-systems within the SM will give us enough hints regarding the
possible stucture of CP violation in and beyond the SM,
and may possibly also
help us to narrow down our searches in that.

In Standard Model, the phenomenon of CP violation
can be established if we can measure
accurately the three angles ($\alpha \equiv \phi_2$, $\beta \equiv
\phi_1$ and $\gamma \equiv \phi_3$) of the CKM unitarity triangle, which
add up to $180^\circ$. The angle $\beta $ is the simplest one and can be
measured in the gold plated mode $B\to J/\psi K_S $ without any
uncertainties \cite{gr89,ln89}.
In fact the value of $ \sin 2\beta $ has recently been reported
\cite{belle01,babar01}. The
angle $\alpha$ can be measured from the decay mode $B\to \pi\pi$
\cite{gr89,ln89}. Although
the presence of penguin contribution introduces some uncertainties,
but with isospin techniques \cite{gr90} the penguin contamination
can be disentangled
so that $\alpha$ can be extracted without hadronic uncertainties.
However, this analysis requires the measurement of $B^0 \to \pi^0 \pi^0$,
which is not feasible with first generation $B$-factory experiments.
Alternatively, $\alpha$ can be determined by applying the isospin
technique to $B\to \rho\pi$ decays \cite{hq11}.
The most difficult to measure is the angle $\gamma$. To this end,
various interesting proposals have been made with a view to obtain
$\gamma$ with lesser or no hadronic uncertainties and the
search is going on to
find out a gold plated mode from which one can extract $\gamma$ cleanly.
It has been shown by
Gronau, London and Wyler (GLW) \cite{ref3}
that $\gamma$ can be extracted
from $B\to DK$ decay rates, employing amplitude triangle relations.
Later, modification with different final states for these decay modes 
has been put forward by Atwood, Dunietz and Soni (ADS)
\cite{ref5} so
that one can hope to find $\gamma$ without uncertainties from the
above decay modes. In Ref. \cite{ref4} the self tagging
modes $B_d^0 \to D^0 K^{*0}$ have been considered for the extraction
of $\gamma$. It has been discussed in Ref. \cite{adk}
that $\gamma$ can also be determined cleanly from
$B_s \to D_s^{\pm}K $ decays. The triangle approach
was also extended to the $B_c$ system \cite{ref11} to extract $\gamma$.
It is shown that the $SU(3)$ relations \cite{ref10}
can be employed between $B\to
K\pi, \pi\pi$ decay amplitudes for the extraction of $\gamma$ and a
nontrivial bound for $\gamma$ can be obtained.
It is also argued that the phase $\gamma$ can be extracted from the
$B_c\to DD_s$ decays \cite{giri01},
without any hadronic uncertainties, since these are
pure tree decay modes and are free from penguin pollutions.
We have recently studied another possibility to find the angle $\gamma$
cleanly from $B_s^0 \to \overline{D^0} \phi$ 
decay modes \cite{giri02}.

As emphasized
above the determination of the angle $\gamma$ is most challenging,
and therefore, it requires further scrutiny.
It should be noted here that the precise
value of $\gamma$ will play a crucial role for testing the validity of CKM
model of CP violation in SM, since the standard technique of the unitarity
triangle will largely depend on it. Hence, the goal is to check all
possible decay modes
and try to determine the angle $\gamma$ as cleanly as possible.
In this paper we would like to show that the angle $\gamma $
can also be extracted from the pure tree decays of the
$\Lambda_b$ baryon i.e, $\Lambda_b \to \Lambda
\{D^0, \overline{D^0}, D^0_{CP}\} $. The advantage
of these decay modes is that these are free from penguin pollutions and
the amplitudes are of similar sizes. Furthermore, neither tagging nor
time dependent studies are required for these decay modes, so $\gamma$
can be extracted cleanly without hadronic uncertainties.
CP violation in the bottom baryon case has earlier been studied by
Dunietz \cite{dn90}. One of us has recently studied the
decay mode $\Lambda_b \to p\pi$ in and beyond SM \cite{rm01}.
Here we would like to emphasize that
the bottom baryon decay modes may serve as alternatives and/or may
supplement the bottom meson decays for the study of CP violation.
Although the branching
ratios of $\Lambda_b $ decay modes are
smaller in comparison to those of the $B$
counterpart modes, but there are certain advantages
for the former case over the later.
The usual problem with the bottom meson (for mixing induced CP
violation) is the tagging and the time evolution of the decaying 
particle are not required for bottom baryon case.
In Ref.\cite{dn90} it has been shown by Dunietz 
that the angle $\gamma$ can be
determined from various baryonic decay modes.
He has analyzed in detail the decay modes
$\Lambda_b \to \Lambda \{D^0, \overline {D^0}, D_{CP}^0\}$ and the
corresponding charge conjugate modes by considering three
specific cases : {\it i.} the $p$-wave dominance,
{\it ii.} the $s$-wave dominance {\it iii.} the $p$- to $s$-wave
ratio to be constant, and has shown that $\gamma $ can be determined
from the partial decay rates of the above six processes. 
In this paper we would like to explore the possibilities of
extracting the angle $\gamma$, without assuming the dominance
of either $s$-wave or $p$-wave. For this purpose we need only the
experimental observables (i.e., the decay rate and angular
distribution parameters) for the above six processes and the information
on $\gamma $ can be extracted from these observables.
The plan of the paper is as follows. 
In section II, we present the phenomenology of hyperon decays.
The formalism of the extraction of the angle
$\gamma $ from the decay modes
$\Lambda_b \to \Lambda \{D^0, \overline {D^0}, D_{CP}^0\}$
is presented in Section III.
Section IV contains our conclusion.

\section{Phenomenology of Hyperon decays}

The study of CP violation in strange hyperon decays has been extensively
studied in Ref. \cite{ref51}, where the phenomenology of hyperon
decays has been discussed in great detail. However for the sake of 
completeness we shall present here the basic features of the
nonleptonic hyperon decays. The most general Lorentz-invariant amplitude
for the decay $\Lambda_b \to \Lambda D^0 $ can be written as

\begin{equation}
{\rm Amp}(\Lambda_b(p_i) \to \Lambda(p_f) D^0(q))=i \bar u_{\Lambda}
(p_f)(A+B \gamma_5)
u_{\Lambda_b}(p_i)\;,\label{eq:r1}
\end{equation}
where $u_{\Lambda} $ and  $u_{\Lambda_b } $ are the Dirac spinors
for $\Lambda $ and $\Lambda_b $ baryons respectively. 
The parameters $A$ and $B$ are complex in general.
The matrix element for the corresponding CP conjugate
process $\bar \Lambda_b \to
\bar \Lambda  \overline{D^0}$ is given as

\begin{equation}
{\rm Amp}(\bar \Lambda_b(p_i) \to \bar \Lambda(p_f) \overline{D^0}(q))
=i \bar v_{\bar\Lambda}
(p_f)(-A^*+B^* \gamma_5)
v_{\bar \Lambda_b}(p_i)\;.\label{eq:r11}
\end{equation}
The spin structure present in Eq. (1) can be easily analyzed by reducing
the $4 \times 4 $ Dirac algebra to the $2 \times 2 $ Pauli algebra.
The matrix element may then be written in terms of S-wave
(parity violating) and P-wave (parity conserving) as

\begin{equation}
{\rm Amp}(\Lambda_b(p_i) \to \Lambda(p_f) D^0(q)) =
\chi_{\Lambda}^\dagger(S+P {\mbox{\boldmath $\sigma $}}
\cdot \hat{\bf q})\chi_{\Lambda_b}\;,\label{eq:e02}
\end{equation}
where $\chi_{\Lambda_b}$ and $\chi_\Lambda $ are the Pauli spinors,
${\mbox {\boldmath $\sigma $}}$ are the Pauli spin matrices
and ${\bf q}$ is the c.o.m. momentum of the final
particles in the rest frame of $\Lambda_b$ baryon.
The  amplitudes S and P are given as

\begin{eqnarray}
S=A \sqrt{\frac{\{(m_{\Lambda_b}+m_\Lambda)^2-m_{D}^2 \}}
{16 \pi m_{\Lambda_b}^2}}
\nonumber\\
P=B \sqrt{\frac{\{(m_{\Lambda_b}-m_\Lambda)^2-m_{D}^2 \}}
{16 \pi m_{\Lambda_b}^2}}\;.\label{eq:r32}
\end{eqnarray}
The experimental observables are the total decay rate 
$\Gamma$ and the asymmetry parameters $\alpha^\prime$, $\beta^\prime$ and 
$\gamma^\prime$ which govern the decay-angular
distribution and the polarization of the final baryon. The decay 
rate is given as
\begin{equation}
\Gamma=2 |{\bf q}|\{|S|^2+|P|^2\}\label{eq:e1}
\end{equation}
and the asymmetry parameters are given as
\begin{eqnarray}
&&\alpha^\prime=\frac{2 {\rm Re} (S^*P)}{|S|^2+|P|^2}\nonumber\\
&&\beta^\prime=\frac{2 {\rm Im} (S^*P)}{|S|^2+|P|^2}\nonumber\\
&& \gamma^\prime = \frac{|S|^2-|P|^2}{|S|^2+|P|^2}\;.\label{eq:e2}
\end{eqnarray}
However, these three angular distribution parameters are not
independent, and are related as
\begin{equation}
{\alpha^\prime}^2+{\beta^\prime}^2+{\gamma^\prime}^2=1\;.
\end{equation}
Thus there are three independent observables ($\Gamma$ and any two of
$\alpha^\prime, \beta^\prime$ and $\gamma^\prime$).
Likewise, the observables for the antihyperon decays ( $\bar \Gamma$, 
$\bar \alpha^\prime$, $\bar \beta^\prime $ and
$\bar \gamma^\prime $) can be written
similar to Eqs. (\ref{eq:e1}) and (\ref{eq:e2}) with $S$ and $P$ replaced by
$\bar S$ and $\bar P$ respectively.

With Eqs. (\ref{eq:e1}) and (\ref{eq:e2}), we can alternatively obtain
the following three observables for nonleptonic hyperon decays
\begin{eqnarray}
&& |S|^2 = \frac{\Gamma}{4|{\bf q}|}\left ( 1+\gamma^\prime \right )
\nonumber\\
&& |P|^2 = \frac{\Gamma}{4|{\bf q}|}\left ( 1-\gamma^\prime \right )
\nonumber\\
&& \tan \Delta = \frac{\beta^\prime}{\alpha^\prime}\;,
\label{eq:e01}
\end{eqnarray}
where $\Delta $ is the relative strong phase between the $S$ and $P$ wave
amplitudes.
After knowing the phenomenology of hyperon decays, we now proceed to
extract the weak phase $\gamma $ from the decay modes
$\Lambda_b \to \Lambda
\{D^0, \overline{D^0}, D_{CP}^0\}$.

\section{Extraction of the angle $\gamma $}

With the advent of hadronic $b$ facilities it becomes possible to
produce $\Lambda_b $ baryons in large numbers. In this paper we show
that the baryonic counterpart of the $B_d^0 \to D^0 K_s $ decay modes
could be well suited to extract the CKM angle
$\gamma $. The corresponding $\Lambda_b$ decays are
$\Lambda_b \to \Lambda \{D^0, \overline{D^0}, D_{CP}^0\} $,
where $D_{CP}^0$ is the CP eigenstate of neutral $D$ meson.
Let us now write the decay amplitudes for the above
processes. Both these amplitudes proceed via the colour suppressed tree
diagrams only and hence are free from penguin pollutions. The amplitude for
$\Lambda_b \to \Lambda D^0$ arises from the quark level transition
$b \to c \bar u s $ and hence has no weak phase in the Wolfenstein
parametrization, while the amplitude $\Lambda_b \to \Lambda \overline
{D^0}$ arises from $b \to u \bar c s $  transition and carries
the weak phase
$e^{-i \gamma}$. The amplitudes also have the strong phases
$e^{i \delta_1^i}$ and $e^{i \delta_2^i}$, where $i=S$ or $P$.
Thus we can write the decay amplitude for $\Lambda_b \to \Lambda D^0 $
process as (here and henceforth we will not write explicitly the
spinors $\chi_{\Lambda_b}$ and $\chi_\Lambda $  in the decay amplitudes)
\begin{equation}
A_1={\rm Amp}(\Lambda_b \to \Lambda D^0)= S_1 e^{i \delta_1^S}
+ P_1 e^{i \delta_1^P} {\mbox{\boldmath  $\sigma $}} \cdot \hat{\bf q}
=  e^{i \delta_1^S}\left (S_1+P_1 e^{i \Delta_1}
{\mbox{\boldmath $\sigma $}} \cdot \hat{\bf q}\right )\;,\label{eq:e3}
\end{equation}
where $S_1$ and $P_1$ are magnitudes of the $S$ and $P$ wave amplitudes
and $\Delta_1=\delta_1^P-\delta_1^S$,
is the relative strong phase between them.
From the decay mode $\Lambda_b \to \Lambda D^0 $, one can extract the
three observables $S_1$, $P_1$ and $\Delta $ using  Eq. (\ref{eq:e01}).
Similarly, one can write the decay amplitude for the process
$\Lambda_b \to \Lambda \overline{D^0}$ as
\begin{equation} 
A_2={\rm Amp}(\Lambda_b \to \Lambda \overline{D^0})
= e^{-i \gamma}\left ( S_2 e^{i \delta_2^S}
+ P_2 e^{i \delta_2^P}{\mbox{\boldmath $\sigma $}} \cdot \hat{\bf q}\right )
= e^{-i \gamma} e^{i \delta_2^S}\left (S_2+P_2 e^{i \Delta_2}
{\mbox{\boldmath $\sigma $}} \cdot \hat{\bf q}\right )\;.
\label{eq:e4}
\end{equation}
Thus from this decay mode we can extract another set of
three observables
$S_2$, $P_2$ and $\Delta_2$.
Now, the amplitudes for the corresponding CP conjugate processes are
given as
\begin{eqnarray}
&&\bar A_1={\rm Amp}(\bar \Lambda_b \to \bar \Lambda
\overline{D^0})= \bar S_1 e^{i \delta_1^S}
+ \bar P_1 e^{i \delta_1^P}{\mbox{\boldmath $ \sigma $}} \cdot \hat{\bf q}
=  e^{i \delta_1^S}\left (\bar S_1+\bar P_1 e^{i \Delta_1}
{\mbox{\boldmath $\sigma $}} \cdot \hat{\bf q}\right )\;,\nonumber\\
&&\bar A_2={\rm Amp}(\bar \Lambda_b \to \bar \Lambda D^0)
= e^{i \gamma}\left (\bar S_2 e^{i \delta_2^S}
+ \bar P_2 e^{i \delta_2^P}{\mbox{\boldmath $\sigma $}}
\cdot \hat{\bf q}
\right )
= e^{i \gamma} e^{i \delta_2^S}\left (\bar S_2+\bar P_2
e^{i \Delta_2}{\mbox{\boldmath $\sigma $}} \cdot \hat{\bf q}
\right )\;,
\label{eq:e9}
\end{eqnarray}
where $\bar S_{1,2}=-S_{1,2}$ and $\bar P_{1,2}=P_{1,2}$. Considering
these two above decay modes, the observables
$\bar S_{1,2}$, $\bar P_{1,2}$ and $\Delta_{1,2}$ can be determined.

We now consider the decay modes $\Lambda_b \to \Lambda D_{\pm}^0$,
where $D_{\pm}^0$ denote the neutral $D$ meson even/odd $CP$
states, defined as $D_\pm^0 =(D^0 \pm \overline{D^0})/\ {\sqrt 2} $.
The $CP$ even state $D_+^0$ can be identified by the $CP$ even decay
products such as $\pi^+ \pi^-$ and $K^+ K^-$, whereas the $CP$
odd state $D_-^0$ can be identified by the $CP$ odd products such as
$K_S \pi^0$, $K_S \rho^0$, $K_S \omega $ and $K_S \phi$.
One can use either of these two
CP eigenstates for the extraction of $\gamma$. Here we are considering
the CP even eigenstate ($D_+^0$), however, the same argument
will also hold for the CP odd state ($D_-^0$).

The amplitude for
$\Lambda_b \to \Lambda D_+^0 $ is thus given as
\begin{equation}
A_+={\rm Amp}(\Lambda_b \to \Lambda D_+^0)= S_+ e^{i \delta_+^S}
+ P_+ e^{i \delta_+^P}{\mbox{\boldmath $\sigma $}} \cdot \hat{\bf q}
=  e^{i \delta_+^S}\left (S_++P_+ e^{i \Delta_+}
{\mbox{\boldmath $\sigma $}} \cdot \hat{\bf q}
\right )\;,\label{eq:e5}
\end{equation}
where $S_+$, $P_+$ are the magnitudes of the $S$ and $P$ wave amplitudes
with phases $e^{i \delta_+^S}$ and $e^{i \delta_+^P}$. It should be
noted that these phases contain both strong and weak components. Thus
from this decay mode we can extract the observables $S_+$, $P_+$ and
$\Delta_+=\delta_+^P-\delta_+^S$.
We can write the amplitude for this decay mode in another form, i.e.,
\begin{eqnarray}
A_+ &=& \frac{1}{\sqrt 2}
\left [{\rm Amp}(\Lambda_b \to \Lambda D^0) +
{\rm Amp}(\Lambda_b \to \Lambda
\overline{D^0})\right ]\nonumber\\
&=& \frac{1}{\sqrt 2}\biggr[ e^{i \delta_1^S}\left (S_1+S_2
e^{i(\sigma_+^S -\gamma)}\right )
+e^{i \delta_1^P}\left (P_1+P_2 e^{i(\sigma_+^P- \gamma)}\right )
{\mbox{\boldmath $\sigma $}} \cdot \hat{\bf q}\biggr]\;,
\label{eq:e6}
\end{eqnarray}
where $\sigma_+^{S,P}=\delta_2^{S,P}-\delta_1^{S,P}$.
Thus comparing Eqs. (\ref{eq:e5}) and (\ref{eq:e6}) we obtain
\begin{eqnarray}
&&S_+ e^{i \delta_+^S}=\frac{1}{\sqrt 2} e^{i \delta_1^S}
\left (S_1+S_2
e^{i(\sigma_+^S -\gamma)}\right )\;,\nonumber\\
&&P_+ e^{i \delta_+^P}=  \frac{1}{\sqrt 2}
e^{i \delta_1^P}\left (P_1+P_2 e^{i(\sigma_+^P- \gamma)}\right )\;.
\label{eq:e8}
\end{eqnarray}
Now the amplitude for the corresponding CP conjugate process, i.e.,
$\bar \Lambda_b \to \bar \Lambda D_+^0 $ is given as

\begin{eqnarray}
\bar A_+ &=&{\rm Amp}(\bar \Lambda_b \to \bar \Lambda D_+^0)=
\frac{1}{\sqrt 2}\left [\bar A_1+\bar A_2 \right ]\nonumber\\
&=&\bar S_+
e^{i \delta_+^{\bar S}}
+ \bar P_+ e^{i \delta_+^{\bar P}}{\mbox{\boldmath $\sigma $}}
\cdot \hat{\bf q}
=  e^{i \delta_+^{\bar S}}\left (\bar S_++\bar P_+
e^{i \bar \Delta_+}{\mbox{\boldmath $\sigma $}}
\cdot \hat{\bf q}\right )\;,\label{eq:e10}
\end{eqnarray}
where $\bar S_+$, $\bar P_+$ are the magnitudes
of the $S$ and $P$ wave amplitudes
with phases $e^{i \delta_+^{\bar S}}$ and $e^{i \delta_+^{\bar P}}$.
The observables obtained from this decay modes are $\bar S_+$, $\bar P_+$
and $\bar \Delta_+ = \bar \delta_+^P-\bar \delta_+^S $.
Substituting the values of $\bar A_1$ and $\bar A_2$ from
Eq. (\ref{eq:e9})
in Eq. (\ref{eq:e10}) we obtain the relations similar
to Eq. (\ref{eq:e8}) as

\begin{eqnarray}
&&\bar S_+ e^{i \delta_+^{\bar S}}=\frac{1}{\sqrt 2} e^{i \delta_1^S}
\left (\bar S_1+\bar S_2
e^{i(\sigma_+^S +\gamma)}\right )\;,\nonumber\\
&&\bar P_+ e^{i \delta_+^{\bar P}}=  \frac{1}{\sqrt 2}
e^{i \delta_1^P}\left (\bar P_1
+\bar P_2 e^{i(\sigma_+^P+ \gamma)}\right )\;.
\label{eq:e11}
\end{eqnarray}
We now use Eqs. (\ref{eq:e8}) and (\ref{eq:e11}) to obtain the weak
phase $\gamma $. To derive the analytic expression for $\gamma$,
we define the combinations of observables
\begin{eqnarray}
&& X = \frac{2 S_+^2 -S_1^2-S_2^2}{2 S_1 S_2}\nonumber\\
&& \bar X =
\frac{2 \bar S_+^2 -\bar S_1^2-\bar S_2^2}{2 \bar S_1 \bar S_2}\label{eq:e12}
\end{eqnarray}
Here we have considered only the $S$ wave components, but similar
combinations can also be derived from the $P$ wave observables and one
can use either set, for the extraction of $\gamma$. Thus, in this method
we need to know only the magnitudes of $S$ and $P$ waves but not the
phase difference between them.

It is now very simple to see that one can obtain an expression 
for $\gamma$ from Eq. (\ref{eq:e12}) as
\begin{equation}
2 \gamma =\arccos \bar X -\arccos X\;,
\end{equation}
with some discrete ambiguities.
One can also obtain  the value of $\sin^2 \gamma$
from Eq. (\ref{eq:e12})  via the relation
\begin{equation}
\sin^2\gamma = \frac{1}{2} \left [1 -
X \bar X \pm \sqrt{(1-X^2)(1- \bar X^2)} \right ]\;.\label{eq:r21}
\end{equation}
One of the sign in Eq. (\ref{eq:r21})
will give the correct value of
$\sin^2 \gamma$, while the other will give the value of the
strong phase $\sin^2 \sigma_+^S $.

One can use both $S$ and $P$ wave observables so that some of the
ambiguities will be reduced and $\gamma $ can be determined cleanly.

Now let us find out the values of the branching ratios for the above decay
modes. The decay processes  $\Lambda_b \to \Lambda D^0 $
and $\Lambda_b \to \Lambda \overline{D^0} $ are governed by the quark
level transitions $b \to c \bar u s $ and $ b \to u \bar c s$
respectively. The effective Hamiltonians for such transitions are given as
\begin{equation}
{\cal H}_{eff}(b \to c \bar u s)= \frac{G_F}{\sqrt 2} V_{cb}V_{us}^*\left [
C_1(m_b) (\bar c b) (\bar su ) + C_2(m_b)(\bar c u)
(\bar s b) \right ]\label{eq:h1}
\end{equation}
and
\begin{equation}
{\cal H}_{eff}(b \to u \bar c s)= \frac{G_F}{\sqrt 2} V_{ub}V_{cs}^*\left [
C_1(m_b) (\bar u b)(\bar sc) + C_2(m_b)(\bar uc)
(\bar s b) \right ]\label{eq:h2}
\end{equation}
respectively, where $C_1(m_b)=1.13 $ and $C_2(m_b)=-0.29 $
are the Wilson coefficients
evaluated at the
renormalization scale $m_b $.  $(\bar cb)=\bar c\gamma^\mu
(1-\gamma_5)b $, etc. are the usual left handed color singlet quark currents.

The hadronic matrix elements of the four-fermion current operators present
in the effective Hamiltonian (\ref{eq:h1}) and (\ref{eq:h2})
are very difficult to evaluate from the first principle of QCD.
So to evaluate the matrix elements of the effective Hamiltonian,
we use the generalized factorization approximation.
In this approximation, the hadronic
matrix elements of the four quark operators split into the product
of two matrix elements, one describing the transition of the initial baryon
to the final baryon state and the other describing the formation of
the final meson from vacuum. Furthermore, the
nonfactorizable contributions, which play an important role in colour
suppressed decays, are incorporated in a phenomenological way :
they are lumped into the coefficients $a_1=C_1+C_2/N_c $ and $a_2
=C_2+C_1/N_c$, so that the effective coefficients
$a_1^{eff}$ and $a_2^{eff}$ are treated as free parameters and their
values can be extracted from the experimental data.
In this paper we shall denote $a_2^{eff}$, which describes the color
suppressed decays, by
simply $a_2$. Thus the factorized amplitude for the decay
process $\Lambda_b \to \Lambda D^0 $ is given as

\begin{equation}
{\rm Amp}(\Lambda_b \to \Lambda D^0)=
\frac{G_F}{\sqrt 2} V_{cb}V_{us}^* a_2 \langle  D^0|
(\bar c u)|0\rangle  \langle \Lambda |
(\bar s b)| \Lambda_b \rangle \;.
\end{equation}
The general expression for
the baryon transition is given as
\begin{eqnarray}
\langle \Lambda (p_f)|V_{\mu}-A_{\mu}|\Lambda_b(p_i) \rangle 
&=&\bar u_\Lambda (p_f) \biggr\{f_1(q^2) \gamma_{\mu} 
+i f_2 (q^2) \sigma_{\mu \nu}
q^{\nu}+f_3 (q^2) q_\mu \nonumber\\
&-&[g_1(q^2) \gamma_{\mu} +i g_2 (q^2) \sigma_{\mu \nu}
q^{\nu}+g_3 (q^2) q_\mu]\gamma_5 \biggr\}u_{\Lambda_b}(p_i)\;,
\end{eqnarray}
where $q=p_i-p_f$.  In order to evaluate the form factors 
at maximum momentum transfer,
we have employed nonrelativistic quark model \cite{ref6}, where
they are given as :
\begin{eqnarray}
\frac{f_1(q_m^2)}{N_{fi}}&=&1-\frac{\Delta m}{2 m_{\Lambda_b}} +
\frac{\Delta m}{4 m_{\Lambda_b}
m_s}\left (1 -\frac{L}{2 m_{\Lambda}}\right )(m_
{\Lambda_b}+m_{\Lambda}- \Delta m)
\nonumber\\
&-&\frac{\Delta m}{8 m_{\Lambda_b} m_\Lambda} 
\frac{L}{m_b}(m_{\Lambda_b}+m_\Lambda- \Delta m)
\nonumber\\
\frac{f_3(q_m^2)}{N_{fi}}&=&\frac{1}{2m_{\Lambda_b}}-\frac{1}
{4 m_{\Lambda_b} m_s}
\left (1 -\frac{L}{2 m_{\Lambda}}\right )
(m_{\Lambda_b}+m_\Lambda- \Delta m)\nonumber\\
&-& \frac{L}{8 m_{\Lambda_b} m_\Lambda m_b}
[(m_{\Lambda_b}+m_\Lambda) +\Delta m]
\nonumber\\
\frac{g_1(q_m^2)}{N_{fi}}&=&1+\frac{\Delta m L}{4}
\left (\frac{1}{m_{\Lambda_b} m_s}-\frac{1}{m_\Lambda m_b} \right )
\nonumber\\
\frac{g_3(q_m^2)}{N_{fi}}&=&-\frac{L}{4}
\left (\frac{1}{m_{\Lambda_b}m_s}+\frac{1}{m_\Lambda m_b} \right )
\end{eqnarray}
where $L = m_\Lambda-m_s$, $\Delta m=m_{\Lambda_b}-m_
\Lambda$,
$q_m^2=\Delta m^2 $, $m_b $ and 
$m_s$ are the constituent quark 
masses of the interacting quarks of initial and final baryons
with values $m_s$=510 MeV and $m_b$=5 GeV.  
$N_{fi} $ is the flavour factor :
\begin{equation}
N_{fi}=_{\rm{flavor~ spin}}\langle \Lambda
|b_s^\dagger b_b|\Lambda_b \rangle
_{\rm{flavor~ spin}}=\frac{1}{\sqrt 3}
\end{equation}
Since the calculation of $q^2$ dependence of form factors 
is beyond the scope of the nonrelativistic quark model 
we will follow the conventional practice to assume a pole 
dominance for the form factor $q^2$ behavior as
\begin{equation}
f(q^2)=\frac{f(0)}{(1-q^2/m_V^2)^2}~~~~~~~~
g(q^2)=\frac{g(0)}{(1-q^2/m_A^2)^2}
\end{equation}
where $m_V(m_A)$ is the pole mass of the vector (axial vector)
meson with the same quantum number as the current under
consideration. The pole masses are taken as $m_V=5.42$ GeV and
$m_A=5.86 $ GeV. Assuming a dipole $q^2$ behavior for form factors, 
and taking the masses of the baryons and $D^0$ meson from Ref. 
\cite{pdg00} we found
\begin{equation}
f_1(m_D^2)=0.08\;,~~m_{\Lambda_b}f_3(m_D^2)=-0.01\;,~~g_1(m_D^2)=0.13
~~{\rm and}~~
m_{\Lambda_b}g_3(m_D^2)=-0.05\;.
\end{equation}
The matrix element $\langle D^0 |A_\mu |0 \rangle $ is related to
the $D^0 $ meson  decay constant $f_D $ as
\begin{equation}
\langle D^0(q) |A_\mu |0 \rangle=-i f_{D} q_{\mu}
\end{equation}
Hence we obtain the transition amplitude for 
$\Lambda_b \to \Lambda D^0$ as
\begin{eqnarray}
A(\Lambda_b  \to   \Lambda D^0 )&=&\frac{G_F}{\sqrt 2}
V_{cb} V_{us}^*i ~f_{D}~a_2~  \bar
u_\Lambda (p_f)
\biggr[
\left (f_1(m_{D}^2)(m_{\Lambda_b}-m_\Lambda)+f_3(m_{D}^2) 
m_{D}^2 \right )\nonumber\\
&+ &
\left (g_1(m_{D}^2)(m_{\Lambda_b}+m_\Lambda)
-g_3(m_{D}^2) m_{D}^2 \right )\gamma_5\biggr]u_{\Lambda_b}(p_i)\;.
\label{eq:r2}
\end{eqnarray}
Comparing the Eqs. (\ref{eq:r1}) and (\ref{eq:r2}) we obtain
the values of the parameters $A$ and $B$ as
\begin{eqnarray}
&&A= \frac{G_F}{\sqrt 2} V_{cb} V_{us}^* a_2  f_D
\left [ f_1(m_D^2) (m_{\Lambda_b}-m_{\Lambda})+f_3(m_D^2) m_D^2
\right ]\nonumber\\
&&B= \frac{G_F}{\sqrt 2} V_{cb} V_{us}^* a_2  f_D
\left [ g_1(m_D^2) (m_{\Lambda_b}+m_{\Lambda})-g_3(m_D^2) m_D^2
\right ]\label{eq:r31}
\end{eqnarray}
Thus with Eqs. (\ref{eq:r31}), (\ref{eq:r32}) and (\ref{eq:e1})
the branching ratio for the process $\Lambda_b
\to \Lambda D^0 $ is given as
\begin{equation}
BR(\Lambda_b \to \Lambda D^0)=4.56 \times 10^{-6}\;,
\end{equation}
where we have used $a_2=0.3$, $f_D=300$ MeV \cite{pdg00} and the values
of the CKM matrix elements from Ref. \cite{pdg00}.
The branching ratio for the decay mode is
$\Lambda_b \to \Lambda \overline{D^0}$ is also found
with Eqs. (\ref{eq:r31}), (\ref{eq:r32}) and (\ref{eq:e1}) and by
substituting the appropriate CKM matrix elements
\begin{equation}                      
BR(\Lambda_b \to \Lambda \overline{D^0})=8.29 \times 10^{-7}\;.
\end{equation}
Although the branching ratios for these decay modes are quite small, 
they could be observed in the future $B$ experiments. 

Now let us study the experimental feasibility of such decay modes.
The observables are the partial decay rate ($\Gamma $) and the
angular distribution parameters $\alpha^\prime$, $\beta^\prime$ and
$\gamma^\prime $, which characterize the strength of $S$ and $P$ waves.
Measuring the partial rate requires no polarization while the measurement of
$\alpha^\prime$, $\beta^\prime$ and $\gamma^\prime $ do. In processes
where the final baryon subsequently decays such as $\Lambda_b \to
\Lambda(\to p \pi^-)D^0 $, $\alpha^\prime$ is determined from the decay
distribution of the final baryon $\Lambda \to p \pi^- $. While the
observables $\beta^\prime$ and $\gamma^\prime $ require an initial
polarization. Here, we would like to point out that because only
a few percent of the $D^0$ modes are $CP$ eigenstates, it might
be difficult to extract the angular distribution parameters
for $\Lambda_b \to \Lambda D_\pm^0 $ modes. The BTeV
experiment, with luminosity $2 \times 10^{32} {\rm cm}^{-2} {s}^{-1}$, will
produce $2 \times 10^{11}$ $b \bar b $ hadrons per $10^7$ sec of running
\cite{ss11}.
If we assume the production fraction as \cite{dun94} 
\begin{equation}
\overline{B_d} : B^- : \overline{B_s} : \Lambda_b
=0.375 : 0.375 : 0.15 : 10
\end{equation}
we get around $2 \times 10^{10}$ numbers of $\Lambda_b $
baryon per year of running
at BTeV. If we take the branching ratios as :
$BR(\Lambda_b \to \Lambda D^0) \sim 4.5 \times 10^{-6}$,
$BR(D^0 \to K^- \pi^+ $
and $ K^- \pi^+ \pi^- \pi^+ )=0.12$, the reconstruction efficiency as
0.05 and the trigger efficiency level as 0.9, we expect to get 
486 $\Lambda D^0 $ events per year.

\section{Conclusion}

In this paper, we have analyzed the possibility of extracting the weak
phase $\gamma$ from the decay modes
$\Lambda_b \to \Lambda\{D^0, \overline{D^0}, D_{CP}^0 \}$.
The transition amplitudes and branching ratios of these decay processes
are calculated using the effective Hamiltonian and the generalized
factorization approximation. The advantage of these modes are that,
first, they are described
by the color suppressed tree diagrams only and are free from penguin
contaminations. Second, neither tagging nor time dependent studies
are required for these decay modes and hence $\gamma $ can be determined
cleanly without any hadronic uncertainties. The price one has to pay is that
the branching ratios for these processes are 
to be very small, of the order $(10^{-6}-10^{-7})$, which are one or two
order smaller than those of the corresponding $B$ decays.
Nevertheless, the absence of usual difficulties, as in the case of
bottom mesons, make such modes worthy of careful study.
To do so, we have to wait
for the future high statistics $B$ experiments, where large samples
of $\Lambda_b$ events are expected to be available.

To summarize, in this paper we have shown that the decay modes
$\Lambda_b \to \Lambda\{D^0, \overline{D^0}, D_{CP}^0 \}$ appear
to be ideally suited for the clean determination of the angle $\gamma $. 
Here we have considered only the Standard Model contributions
to the decay processes. So if the extracted value of $\gamma $, from these
modes differs from its value constrained by SM, then this would be a clear
indication of the possible existence of new physics. Therefore,
outside the $B$ meson systm, these
decay modes may possibly give us valuable information regarding the
nature of CP violation and guide us to know physics beyond Standard Model.
The strategies presented here are particularly interesting for future
$B$ experiments such as BTeV, LHCb and beyond.

\acknowledgements
We are thankful to Professor N. G. Deshpande for valuable discussions.
AKG and MPK would like to thank Council of Scientific and Industrial
Research, Government of India, for financial support.


\begin{thebibliography}{99}
\bibitem{kog89} {\it CP Violation} ed. C. Jarlskog (World Scientific,
Singapore, 1989)
\bibitem{bigi00} {\it CP Violation} by I. I. Bigi and A. I. Sanda,
Cambridge Monographs on Particle Physics, Nuclear Physics and
Cosmology, 2000).
\bibitem{lav00} {\it CP Violation} by  G. C. Branco, L. Lavoura
and J. P. Silva, International
series on Monographs on Physics, Number 103, Oxford University Press (1999).
\bibitem{ref1} N. Cabibbo, Phys. Rev. Lett. {\bf 10}, 531 (1963);
M. Kobayashi and T. Maskawa, Prog. Theo. Phys.
{\bf 49}, 652 (1973).
\bibitem{buras98} A. J. Buras and R. Fleischer, in {\it Heavy Flavours
II} eds. A. J. Buras and M. Lindner (World Scientific, Singapore, 1998)
p. 65.
\bibitem{quin98} {\it The BaBar Physics Book} eds. P. F. Harrison and H. R.
Quinn (SLAC report 504, 1998).
\bibitem{chau94} L. L. Chau and W.-Y. Keung, Phys. Rev. Lett. {\bf 53},
1802 (1984); C. Jarlskog and R. Stora, Phys. Lett. {\bf B 208}, 268 (1988);
G. C. Branco and L. Lavoura, Phys. Lett. {\bf B 208}, 123 (1988).
\bibitem{stone94} For a review see {\it B Decays} ed. S. Stone
(World Scientific, Singapore 1994).
\bibitem{belle01} A. Abashian et al, The BELLE Collaboration, Phys. Rev.
Lett. {\bf 86}, 2509 (2001); K. Abe et al, Phys. Rev. Lett. {\bf 87},
091802 (2001).
\bibitem{babar01} B. Aubert et al, BABAR Collaboration, Phys. Rev. Lett.
{\bf 86}, 2515 (2001); {\it ibid} {\bf 87}, 091801 (2001).
\bibitem{gr89} M. Gronau, Phys. Rev. Lett. {\bf 63}, 1451 (1989).
\bibitem{ln89} D. London and R. D. Peccei, Phys. Lett.
{\bf B 223}, 257 (1989).
\bibitem{gr90} M. Gronau and D. London, Phys. Rev. Lett {\bf 65},
3381 (1990).
\bibitem{hq11} A. E. Snyder and H. R. Quinn, Phys. Rev. {\bf D 48 }, 2139 (1993).
\bibitem{ref3} M. Gronau and D. London, Phys. Lett. {\bf B 253}, 483(1991):
M. Gronau and D. Wyler, Phys. Lett. {\bf B 265}, 172 (1991).
\bibitem{ref5} D. Atwood, I. Dunietz and A. Soni, Phys. Rev. Lett
{\bf 78}, 3257 (1997); Phys. Rev. {\bf D 63}, 036005 (2001).
\bibitem{ref4} I. Dunietz, Phys. Lett. {B 270}, 75 (1991).
\bibitem{adk} R. Aleksan, I Dunietz and B. Kayser, Zeit. Phys. {\bf C 54}, 653 (1992).
\bibitem{ref11} R. Fleischer and D. Wyler, Phys. Rev. {\bf D 62}, 057503
(2000).
\bibitem{ref10} M. Gronau, J. L. Rosner and D. London, Phys. Rev.
Lett. {\bf 73}, 21 (1994); R. Fleischer, Phys. Lett. {\bf B 365}, 399 (1996);
M. Neubert and J. L. Rosner, Phys. Rev. Lett. {\bf 81}, 5076 (1998).
\bibitem{giri01} A. K. Giri, R. Mohanta and M. P. Khanna, Phys. Rev.
{\bf D 65}, 034016 (2002).
\bibitem{giri02} A. K. Giri, R. Mohanta and M. P. Khanna, hep-ph/0112016,
to appear in Phys. Rev D.
\bibitem{dn90} I. Dunietz, Z. Phys. {\bf C 56}, 129 (1992).
\bibitem{rm01} R. Mohanta, Euro. Phys. Jour. {\bf C 16}, 289 (2000);
Phys. Rev. {\bf D 63}, 056006 (2001).
\bibitem{ref51} J. F. Donoghue and S. Pakvasa, 
Phys. Rev. Lett {\bf 55}, 162 (1985); J. F. Donoghue, X-G. He 
and S. Pakvasa, Phys. Rev. {\bf D 34}, 833 (1986); 
D. Chang, X-G. He and S. Pakvasa, Phys. Rev. Lett.
{\bf 74}, 3927 (1995); X-G. He, H. Steger and G. Valencia, Phys.
Lett. {\bf B 272}, 411 (1991); 
X-G. He and G. Valencia, Phys. Rev. {\bf D 52}, 5257 (1995); 
J. Tandean and G. Valencia, Phys. Lett. {\bf 451}, 382 (1999).
\bibitem{ref6} H. Y. Cheng and B. Tseng, Phys. Rev. {\bf D 53}, 1457 (1996);
H.-Y. Cheng, Phys. Rev. {\bf D 56}, 2799 (1997).
\bibitem{pdg00} D. E. Groom et al, Review of Particle Physics, The
Euro. Phys. Jour. {\bf C 15}, 1 (2000).
\bibitem{ss11} S. Stone, in "Proc. of Third International Conference on B Physics 
and CP violation" Ed. H.-Y. Cheng and W.-S. Hou, World Scientific,
(Singapore, 2000) p. 450 (hep-ph/0002025).
\bibitem{dun94} I. Dunietz, FERMILAB-PUB-163-T (hep-ph/9409355).
\end{thebibliography}
\end{document}